# Compressive-strain induced enhancement of exchange interactions and short-range magnetic order in $Sr_2IrO_4$ investigated by Raman spectroscopy

A. Seo,[1,2] P. P. Stavropoulos,[3] H.-H. Kim,[2] K. Fürsich,[2] M. Souri,[1] J. G. Connell,[1] H. Gretarsson,[2] M. Minola,[2] H. Y. Kee,[3] and B. Keimer[2]

[1.] *Department of Physics and Astronomy, University of Kentucky, 500 Rose St., Lexington, Kentucky 40506, USA*
[2.] *Max-Planck-Institut für Festkörperforschung, Heisenbergstraße 1, D-70569 Stuttgart, Germany*
[3.] *Department of Physics, University of Toronto, 60 St. George St., Toronto, Ontario, Canada M5S 1A7*

**Abstract**

We have carried out Raman spectroscopy experiments to investigate two-magnon excitations in epitaxial thin films of the quasi-two-dimensional antiferromagnetic Mott insulator $Sr_2IrO_4$ under in-plane misfit strain. With in-plane biaxial compression, the energy of the two-magnon peak increases, and the peak remains observable over a wider temperature range above the Néel temperature, indicating strain-induced enhancement of the superexchange interactions between $J_{eff} = 1/2$ pseudospins. From density functional theory calculations, we have found an increase of the nearest-neighbor hopping parameter and exchange interaction with increasing biaxial compressive strain, in agreement with the experimental observations. Our experimental and theoretical results provide perspectives for systematic, theory-guided strain control of the primary exchange interactions in $5d$ transition metal oxides.

The coupling between different interactions with similar energy scales leads to new ground states in solids. Among many examples, the interplay between the relativistic spin-orbit interaction ($\lambda$) and the on-site Coulomb interaction ($U$) in materials with strongly correlated electrons has attracted great attention. Depending on their energy scales, the confluence of these two interactions is expected to result in novel phases such as Weyl semimetals, axion insulators, topological Mott insulators, etc. [1]. Indeed, spin-orbit coupled Mott insulators with $J_{eff} = 1/2$ pseudospins have been discovered in $5d$ systems such as the layered iridates $Sr_2IrO_4$ [2] and $Na_2IrO_3$ [3] where the energy scales of $\lambda$ and $U$ are comparable.

The $J_{eff} = 1/2$ pseudospins and their exchange interactions have been considered as analogs of spin $S = 1/2$ states. The mapping of the $J_{eff} = 1/2$ pseudospin into $S = 1/2$ on a square lattice makes the quasi-two-dimensional (quasi-2D) magnetism in $Sr_2IrO_4$ closely comparable to $La_2CuO_4$, which is a parent compound of the high-temperature superconducting cuprates. Recent theoretical and experimental studies have focused on the similarities of both systems [4-7]. However, while the $S = 1/2$ electrons of the $Cu^{2+}$ ions reside in non-degenerate $e_g$ orbitals and therefore exhibit weak coupling with the lattice, the $J_{eff} = 1/2$ pseudospin wave function of $Ir^{4+}$ ions in iridates is an entangled state of both spin-up and spin-down states and all $t_{2g}$ orbitals [2,8]. Therefore, multiple $d$-orbitals must be considered for a realistic description of the layered iridates, in analogy to multi-orbital systems such as the manganates and in contrast to the orbitally non-degenerate cuprates. Note that the multi-spin/orbital characteristic can provide an intriguing scheme of enabling the manipulation of fundamental interactions via external parameters. For example, one could tune the magnetic exchange interaction of the layered iridates by applying external strain to its lattice. In fact, the exchange interaction plays an important role not only in magnetism but also in novel electronic properties. For instance, short-range antiferromagnetic correlations and spin fluctuations, which are governed primarily by nearest-neighbor superexchange interactions, are thought to be key to understanding the pairing mechanisms of unconventional superconductors such as the cuprates, Fe-pnictides, chalcogenides, and some heavy-fermion materials [9]. Hence, understanding how the primary exchange interaction is affected by external perturbations is essential for elucidating the origin of the macroscopic properties of many strongly coupled materials.

In this manuscript, we report Raman spectroscopic studies on $Sr_2IrO_4$ epitaxial thin films under different degrees of misfit strain induced by the substrate to reveal the interactions between the $J_{eff} = 1/2$ pseudospins and the lattice. We have observed clear two-magnon peaks, which contain direct information on the exchange interactions between the $J_{eff} = 1/2$ pseudospins in $Sr_2IrO_4$. As in-plane compressive strain is applied, the two-magnon peak is shifted to higher energies, indicating that the exchange interaction is strengthened. Moreover, the in-plane compressive strain enhances the 2D short-range magnetic order that persists at temperatures above the onset of three dimensional long-range order. Our results thus demonstrate the substantial influence of lattice strain on the exchange interaction, which should be carefully considered to understand the physics of strongly interacting, multi-orbital systems.

**Table I.** Lattice parameters and misfit strain values of $Sr_2IrO_4$ thin films (~50 nm thick).

| Samples | $a$ (Å) | $c/2$ (Å) | Misfit strain tensile (+) / compressive (-) | |
| --- | --- | --- | --- | --- |
| | | | In-plane (%) | Out-of-plane (%) |
| $Sr_2IrO_4$/STO | 3.90 | 12.83 | +0.4 (±0.1) | -0.50 (±0.1) |
| $Sr_2IrO_4$/LSAT | 3.86 | 12.93 | -0.7 (±0.2) | +0.32 (±0.1) |

The Sr$_2$IrO$_4$ epitaxial thin-films were grown by pulsed laser deposition (PLD) on various oxide single-crystal substrates, *i.e.* SrTiO$_3$ (001) (STO), (LaAlO$_3$)$_{0.3}$(Sr$_2$TaAlO$_6$)$_{0.7}$ (001) (LSAT), NdGaO$_3$ (110) (NGO), and LaAlO$_3$ (110) (LAO), with the orientations of the *c*-axis perpendicular to the substrate surface. We have selected these substrates because their in-plane lattice mismatches ($\equiv \frac{a_{substrate}-a_{bulk}}{a_{substrate}} \times 100\ \%$) with bulk single crystal Sr$_2$IrO$_4$ provide a wide range of values, *i.e.* +0.4 % (STO), -0.5 % (LSAT), -2.0 % (NGO), and -2.6 % (LAO), thereby covering both tensile (positive sign) and compressive (negative sign) strain. The thickness of the Sr$_2$IrO$_4$ thin films is kept at ~50 nm, in order to obtain enough Raman cross-section from the thin films while avoiding major relaxation of the misfit strain. The details of the PLD conditions are described in Refs. [10,11]. From X-ray diffraction reciprocal space mapping (Fig. 1 (a) and Fig. S1), the volume-averaged strain values ($\equiv \frac{a_{film}-a_{bulk}}{a_{bulk}} \times 100\ \%$) are estimated about +0.4 % (STO), -0.7 % (LSAT), -1.9 % (NGO), and -1.2 % (LAO), as summarized in Table I. The thin films grown on NGO and LAO substrates exhibit incoherent lattice strain relaxation (Fig. S1) presumably due to relatively large in-plane lattice mismatches between the films and the substrates. Hence, here we focus on the Sr$_2$IrO$_4$ thin films grown on STO and LSAT substrates. Since the thermal expansion coefficients of these oxide substrates are known to be similar (~10$^{-5}$ K$^{-1}$) [12], we assume that the lattice strain is essentially maintained at low temperatures. Compared to the hydrostatic pressure dependent study of Sr$_2$IrO$_4$ single crystals [13], note that the epitaxial strain here is less than about 2 GPa in pressure. Hence, it is a good tuning parameter for studying the interactions between the $J_{eff}$ = 1/2 pseudospins since the intermixing of the $J_{eff}$ =1/2 and 3/2 states is expected at much higher pressure.

The Raman spectra of the Sr$_2$IrO$_4$ thin films were obtained by using a confocal micro-Raman (JobinYvon LabRam HR800) spectrometer with a focused beam spot size of ~5 μm, as shown schematically in Fig. S2 (a), with the 632.8 nm (1.96 eV) excitation line of a HeNe laser and a 600 grooves/mm grating with energy resolution ~5 cm$^{-1}$. Note that the photon energy of the laser, 1.96 eV, is near the charge-transfer gap (from O 2$p$ to Ir 5$d$ bands) of Sr$_2$IrO$_4$ [14], thereby resulting in effective resonant Raman scattering. The power of the laser at the sample surface is kept below ~0.8 mW to minimize laser heating effects. The temperature at the sample was confirmed by measuring the intensity ratio of low-energy phonon modes of the substrates in Stokes and Anti-Stokes regimes. Since the mass of the substrates is about 10$^4$ times larger than those of the thin films and the optical gap energies of the substrates (> 3 eV) are much higher than the 1.96 eV excitation energy, it is reasonable to assume that the thin films are in thermal equilibrium with the substrates at each temperature. The Raman spectra of the thin films are extracted by subtracting the substrates' contributions from the raw spectra measured with an optimally focused beam and the confocal setup shown in Fig. S2 (b), as described in Refs. [15,16]. Figure S2 (c) shows that the $A_{1g}+B_{1g}$, $B_{2g}$, $A_{1g}+B_{2g}$, and $B_{1g}$ phonon modes are observed in the polarization-dependent backscattering channels $z(xx)\bar{z}$, $z(xy)\bar{z}$, $z(x'x')\bar{z}$, and $z(x'y')\bar{z}$, respectively, where the axes $x'$ and $y'$ are along the Ir-O bond direction and the axes $x$ and $y$ are rotated 45° from the bond [17]. These polarization dependent phonon selection rules are consistent with those of Sr$_2$IrO$_4$ single crystals, as reported in Ref. [18].

We checked if there is any evidence of oxygen vacancies, of which precise control is a formidable task in the synthesis of complex oxide thin films. It is known that a new phonon mode at ~258 cm$^{-1}$ (32 meV) in $z(xx)\bar{z}$ polarization emerges in oxygen-deficient Sr$_2$IrO$_4$ [19]. However, this oxygen-vacancy mode ($V_O$) is absent in the Raman spectra of our thin-film samples, as shown in Fig. 1 (b), indicating that the density of oxygen vacancies in our Sr$_2$IrO$_4$ thin films remains negligible owing to the relatively high oxygen partial pressure ($P_{O_2}$ = 10 mTorr) used during the thin-film synthesis and post-annealing procedures.

We have observed broad $B_{2g}$ peaks originating from two-magnon scattering at ~1320 cm$^{-1}$ in all of the Sr$_2$IrO$_4$ thin films at low temperatures, as shown in Fig. 2 (a) and Fig. S3 (a). As the temperature increases, the spectral weight of the two-magnon peak weakens due to magnon-magnon interactions. The overall spectral shape of the two-magnon scattering is consistent with that observed in Sr$_2$IrO$_4$ single crystals [18,20]. Note that such a clear two-magnon peak is not always seen in Sr$_2$IrO$_4$ bulk crystals, particularly in off-stoichiometric samples, as discussed in Refs. [21,22]. Hence, our results confirm that these Sr$_2$IrO$_4$ thin film samples are free from extensive off-stoichiometry, regardless of the misfit strain. The two-magnon peak energy allows us to directly estimate the exchange interaction ($J$) of the $J_{eff}$ = 1/2 pseudospins in the Sr$_2$IrO$_4$ thin films. According to the two-dimensional (2D) Heisenberg model utilizing $S$ = 1/2 with quantum fluctuations, the two-magnon peak energy is expected to be approximately at 2.7$J$ [23], which is reduced from the classical prediction of 3$J$ [24]. Since the nearest-neighbor $J$ of Sr$_2$IrO$_4$ is known to be ~60 meV [25], the observed two-magnon peak at ~1320 cm$^{-1}$ (~164 meV, implying $J \approx$ 60 meV) is very reasonable, as also discussed for Sr$_2$IrO$_4$ single crystals in Ref. [18]. However, the asymmetric shape of the two-magnon scattering is not quantitatively understood, preventing fitting to concrete model calculations at this moment. There are some ideas including a recent theoretical work [26] suggesting that the asymmetric shape might arise from amplitude fluctuations of the magnetic order parameter, the so-called Higgs mode. Alternatively, longer-range exchange interactions may also affect the shape of the two-magnon feature [19]. Regardless of details of the line shape, we have determined the two-magnon peak energies ($\omega_{2M}$) through fits to a model function comprising two Lorentz oscillators, with imaginary part $\chi''(\omega) = \chi''_1(\omega) + \chi''_2(\omega) = \sum_{n=1}^{2} \frac{a_n \cdot \Gamma_n \cdot \omega_{2M_n} \cdot \omega}{(\omega_{2M_n}^2 - \omega^2)^2 + \Gamma_n^2 \cdot \omega^2}$, where $a_n$ and $\Gamma_n$ are the amplitude and the width of the *n*-th mode, respectively. This phenomenological function fits the experimental spectra very well, as shown in Fig. 2 (b).

Note that the two-magnon peaks ($\omega_{2M_1}$) shift to higher energies as compressive strain is applied to the Sr$_2$IrO$_4$ thin films. To confirm the two-magnon Raman scattering peak shift, we have calculated the integrated intensities of the raw spectra, *i.e.* the spectral weights, as a function of Raman shift. The result also shows clear blueshift (Fig. S4). Hence, the observation of the two-magnon peak shift does not depend on details of the curve fitting, but should be regarded as a robust experimental result. Figure 3 (a) summarizes the two-magnon peak energies ($\omega_{2M_1}$) as a function of temperature. The shift of the peak maximum from Sr$_2$IrO$_4$/STO (+0.4 % tensile strain) to Sr$_2$IrO$_4$/LSAT (-0.7 % compressive strain) is approximately 50 cm$^{-1}$ (6.2 meV) at 10 K. At higher temperatures than 10 K, a similar increase of $\omega_{2M}$ values is also observed in Sr$_2$IrO$_4$/LSAT compared to Sr$_2$IrO$_4$/STO, until the two-magnon peak becomes very broad near $T_N$, as indicated in the increase of $\Gamma_1$ upon heating (Fig. 3 (b)). This experimental observation indicates that the overall 1 % change of the in-plane compressive strain enhances *J* by about 4 % ($\Delta J \approx 2.2$ meV), *i.e.* from *J* = 61.1 meV (Sr$_2$IrO$_4$/STO) to *J* = 63.2 meV (Sr$_2$IrO$_4$/LSAT). From *dc*-magnetometry measurements using a superconducting quantum interference device, we have observed that the Néel temperature ($T_N$) also increases about 4 % (~10 K) due to the increased compressive strain from Sr$_2$IrO$_4$/STO to Sr$_2$IrO$_4$/LSAT, as shown in the inset of Fig. 3 (b) [27]. The increased $T_N$ is consistent with the increased *J* since both parameters should be monotonically related in antiferromagnetic systems.

It is tempting to explain these observations with a simple magneto-striction effect. However, the increase of $J$ with compressive strain contradicts the widely held picture that metal oxides with perovskite structure respond to strain primarily via tilts and rotations of rigid metal-oxide octahedra, as in the Glazer description [28]. The hopping integral ($t$) is proportional to $\frac{\sin\theta/2}{d_{Ir-O}^{3.5}}$, where $\theta$ is the bond angle of Ir-O-Ir and $d_{Ir-O}$ is the bond length between Ir and O ions in the IrO$_2$ planes. The modest misfit strain of less than a few percent is expected to reduce $\theta$ rather than $d_{Ir-O}$ according to the rigid octahedral picture. Hence, compressive misfit strain should not increase but decrease $t$ and $J$ since $J \propto \frac{t^2}{\widetilde{U}}$, where $\widetilde{U}$ is the effective on-site Coulomb interaction. However, our experimental data clearly demonstrate the opposite scenario, *i.e.* $J$ increases as compressive strain is applied to Sr$_2$IrO$_4$ thin films.

To understand our observations, we have performed *ab initio* calculations and found that the effective Heisenberg interaction of the $J_{\text{eff}} = 1/2$ pseudospins increases as the in-plane compressive strain increases. We have employed the Vienna *ab initio* simulation package (VASP) code to relax the ionic positions of Sr$_2$IrO$_4$ under the constraint of different in-plane lattice strains. The details of the nearest and second nearest neighbor hopping parameters of the $t_{2g}$ orbitals using the optimized lattice constants and $\theta$ are described in the Supplemental Material [17]. As expected, the hopping integral between $d_{xy}$-orbitals ($t_{xy}$) slightly decreases under compressive strain due to the decrease of $\theta$. However, the compressive strain enhances the $d_{xz}$ ($d_{yz}$) hopping integral, $t_{xz}$ ($t_{yz}$), along the $x$ ($y$)-axis. Therefore, the overall hopping integral $t = \frac{1}{\sqrt{3}}(t_{xy} + t_{yz} + t_{xz})$ of the $J_{\text{eff}} = 1/2$ pseudospins increases under compressive strain. Furthermore, due to the staggered rotation of the IrO$_6$ octahedra, the inter-orbital hopping $t_z = \frac{1}{\sqrt{2}}(t_{xz,yz} - t_{yz,xz})$ [7], where $t_{xz,yz}$ and $t_{yz,xz}$ are the off-diagonal terms in the $t_{2g}$ matrix, also increases under the compressive strain. Projecting the $t_{2g}$ orbitals onto the $J_{eff} = 1/2$ manifold and taking the large Hubbard $U$ limit leads to an antiferromagnetic Heisenberg model with $J = \sqrt{J_1^2 + D^2}$, where $J_1 = 4\frac{t^2 - t_z^2}{U}$ and $D$ represents the Dzyaloshinskii-Moriya (DM) interaction proportional to $\frac{8tt_z}{U}$ [7]. The effective SU(2) Heisenberg model is obtained because $t_z$ can be absorbed into $t$ by a unitary transformation [4]. The results are summarized in Table II where $\tilde{U} = U - J_H = 2$ eV and $J_H$ is the Hund's coupling. Note that both $J_1$ and $D$ increase because the effective hopping integrals of the $J_{eff} = 1/2$ pseudospins originate from a mixture of $t_{2g}$ orbitals and $\theta$ has opposite effects on the $t_{2g}$ orbitals. The second-nearest neighbor Heisenberg interaction ($J_2$) is of the order of 1 meV, so we ignored its contribution to the magnon energy [29]. The calculations show qualitatively consistent results with our experiments.

**Table II.** Effective hopping parameters $t$ and $t_z$, Heisenberg, and DM interactions for $J_{eff} = 1/2$ pseudospin obtained for various in-plane lattice strain.

|  | Pseudo-cubic in-plane (Å) | $t$ (meV) | $t_z$ (meV) | $J$ (meV) | $J_1$ (meV) | $\lvert D \rvert$ (meV) |
|---|---|---|---|---|---|---|
| Sr$_2$IrO$_4$/STO | 3.90 | 157 | 19 | 50 | 49 | 12 |
| Sr$_2$IrO$_4$/LSAT | 3.87 | 159 | 17 | 51 | 50 | 11 |
| Sr$_2$IrO$_4$/LAO | 3.79 | 171 | 20 | 60 | 58 | 14 |

The increased $t$ under compressive strain also gives an answer to a puzzling observation of an optical spectroscopic study, namely the shift of the optical gap of magnitude ~0.3 eV to lower energies under compressive strain [10]. Since the low-energy optical absorption is understood as an optical transition from the lower Hubbard band to the upper Hubbard band of the $J_{eff} = 1/2$ state, *i.e.* the Mott-Hubbard gap, its energy is proportional to $U/t$ [14,30]. This observation has been debated because compressive strain is expected to decrease $\theta$ and $t$, as generally believed in the rigid octahedral picture of transition metal oxides. However, if compressive strain enhances $t$ due to the multi-orbital character of the $J_{eff} = 1/2$ pseudospins in this layered iridate, the red-shift of the Mott-Hubbard gap can thus be understood consistently with the reduced $U/t$ and the increased two-magnon peak energy observed in this study.

It is also noteworthy that the spectral intensities of the two-magnon peaks at $T > T_N$ for the samples under compressive strain, *i.e.* $Sr_2IrO_4$/LSAT, $Sr_2IrO_4$/NGO, and $Sr_2IrO_4$/LAO, are remarkably higher than the tensile-strained $Sr_2IrO_4$/STO due to enhanced short-range magnetic order in this quasi-2D system. The two-magnon density of states is dominated by zone-boundary excitations, which can be approximated as local spin flips, reflecting quasi-2D spin-spin correlations. Hence, the two-magnon peak intensities are often observed at temperatures above the onset of long-range magnetic order, *i.e.* above $T_N$, as known for $La_2CuO_4$ [31]. Indeed, Figure 2 (a) shows that the two-magnon peaks have non-zero intensities at temperatures above $T_N$. The integrated two-magnon intensities from 900 cm$^{-1}$ to 2500 cm$^{-1}$ are plotted as a function of temperature in Fig. 3 (c). Whereas $Sr_2IrO_4$/STO with tensile strain shows little two-magnon scattering intensity at room temperature (300 K), $Sr_2IrO_4$/LSAT with compressive strain exhibits clear non-zero two-magnon peak intensities even at 320 K, which is about 80 K higher than $T_N$. It is interesting to consider why the difference of $T_N$ between $Sr_2IrO_4$/LSAT and $Sr_2IrO_4$/STO is only about 10 K whereas the short-range magnetic order, observed by the two-magnon peak intensities, survives at much higher temperatures for the compressively strained sample. In this quasi-2D antiferromagnetic system, long-range magnetic ordering is governed by small but non-negligible interlayer interactions, which cannot be measured by our current approach. However, it is reasonable to assume that the interlayer interactions are inversely proportional to the interlayer distance, which can be modified by in-plane strain in the hypothesis of elastic deformation. As summarized in Table I, the tensile strain in Sr$_2$IrO$_4$/STO reduces the $c$-axis lattice parameter of the Sr$_2$IrO$_4$ thin film while the compressive strain in Sr$_2$IrO$_4$/LSAT causes relatively small elongation along the $c$-axis. This is presumably because the thermodynamically stable phase of Sr$_2$IrO$_4$ has somewhat elongated IrO$_6$ octahedra with a tetragonal distortion, preventing a further increase in the $c$-axis lattice parameters. The effect of in-plane tensile strain, which reduces $J$ within the quasi-2D layers, on $T_N$ may thus be partially compensated by an enhanced interlayer coupling, thus reducing the overall effect.

Finally, we discuss possible extrinsic factors such as defects other than lattice strain that can affect the experimental observations above. One may argue that the observed change of the two-magnon peak as a function of misfit strain might be caused by different oxygen stoichiometry in these thin-film samples. However, as discussed earlier, the oxygen-vacancy mode at ~258 cm$^{-1}$ (32 meV) [19] is not observed in the Raman spectra of $z(xx)\bar{z}$ polarization (Fig. 1 (b)). Hence, it is reasonable to assume that the density of oxygen vacancies is negligible in our Sr$_2$IrO$_4$ thin films. Therefore, we exclude oxygen deficiencies as a primary reason for the changes in the two-magnon peaks in these Sr$_2$IrO$_4$ thin films.

In conclusion, our Raman spectroscopic studies show that the magnitude of $J$ and the short-range magnetic order of the $J_{\text{eff}} = 1/2$ pseudo-spins in Sr$_2$IrO$_4$ can be modified by lattice strain. Our observations seem somewhat counterintuitive at first glance: whereas small compressive strain (less than a few percent) is considered to decrease the metal-oxygen-metal bond angles and $t$ between neighboring sites in many transition metal oxides, our experimental and theoretical results on Sr$_2$IrO$_4$ thin films demonstrate that compressive strain enhances $J$ and $t$ between the Ir sites. We find that this unexpected behavior is related to the multi-orbital characteristics of the $J_{eff} = 1/2$ wave functions in this layered iridate, which is distinct from orbitally non-degenerate $S = 1/2$ systems of other transition-metal oxides. Hence, when the $J_{eff} = 1/2$ pseudospin is mapped into $S = 1/2$, its multi-orbital nature and coupling with lattice should be considered carefully. Detailed analysis of temperature-dependent phonon modes using high-resolution Raman spectroscopy will also be useful to clarify this issue. However, extracting thin-film phonon spectra is difficult because the raw spectra contain significant contributions of sharp modes from the substrate at low temperatures, and is therefore left for future work.

Our results also suggest some interesting questions and perspectives. For example, to what extent will the lattice strain be effective in enhancing the exchange interaction of the pseudospins? In other words, is it possible to quench the orbital dynamics by large strain so that new electronic and magnetic ground states could emerge via strong spin-orbit coupling? Our current approach of epitaxial growth of thin films is rather limited for this purpose since strain relaxation is expected for higher lattice mismatch. However, the combination of Raman and X-ray spectroscopies with external hydrostatic and/or uniaxial pressure in future studies could shed further light on the coupling between the lattice and pseudospins of layered iridates.

**Acknowledgement**

We thank J. Yu, G. Khaliullin, J. Kim, B. J. Kim, and E. M. Pärschke for fruitful discussions. The sample synthesis and characterizations were supported by National Science Foundation Grant No. DMR-1454200. A.S. acknowledges the support (Research Fellowship for Experienced Researchers) from the Alexander von Humboldt Foundation. B.K. acknowledges financial support from the Deutsche Forschungsgemeinschaft (DFG, German Research

Foundation) – Projektnummer 107745057 - TRR 80. The theoretical work was supported by the Natural Sciences and Engineering Research Council of Canada and the Center for Quantum Materials at the University of Toronto (P.P.S., H.Y.K.). The computations were performed on the GPC and Niagara supercomputers at the SciNet HPC Consortium. SciNet is funded by: the Canada Foundation for Innovation under the auspices of Compute Canada; the Government of Ontario; Ontario Research Fund - Research Excellence; and the University of Toronto.**References**

[1]     W. Witczak-Krempa, G. Chen, Y. B. Kim, and L. Balents, Annu. Rev. Condens. Matter Phys. **5**, 57 (2014).

[2]     B. J. Kim *et al.*, Phys. Rev. Lett. **101**, 076402 (2008).

[3]     R. Comin *et al.*, Phys. Rev. Lett. **109**, 266406 (2012).

[4]     F. Wang and T. Senthil, Phys. Rev. Lett. **106**, 136402 (2011).

[5]     H. Watanabe, T. Shirakawa, and S. Yunoki, Phys. Rev. Lett. **110**, 027002 (2013).

[6]     Y. K. Kim, O. Krupin, J. D. Denlinger, A. Bostwick, E. Rotenberg, Q. Zhao, J. F. Mitchell, J. W. Allen, and B. J. Kim, Science **345**, 187 (2014).

[7]     J.-M. Carter, V. Shankar V., and H.-Y. Kee, Phys. Rev. B **88**, 035111 (2013).

[8]     G. Jackeli and G. Khaliullin, Phys. Rev. Lett. **102**, 017205 (2009).

[9]     D. J. Scalapino, Rev. Mod. Phys. **84**, 1383 (2012).

[10]    J. Nichols, J. Terzic, E. G. Bittle, O. B. Korneta, L. E. D. Long, J. W. Brill, G. Cao, and S. S. A. Seo, Appl. Phys. Lett. **102**, 141908 (2013).

[11]    S. S. A. Seo, J. Nichols, J. Hwang, J. Terzic, J. H. Gruenewald, M. Souri, J. Thompson, J. G. Connell, and G. Cao, Appl. Phys. Lett. **109**, 201901 (2016).

[12]    A. Okazaki and M. Kawaminami, Mater. Res. Bull. **8**, 545 (1973).

[13]    D. Haskel, G. Fabbris, M. Zhernenkov, P. P. Kong, C. Q. Jin, G. Cao, and M. van Veenendaal, Phys. Rev. Lett. **109**, 027204 (2012).

[14]    C. H. Sohn, M.-C. Lee, H. J. Park, K. J. Noh, H. K. Yoo, S. J. Moon, K. W. Kim, T. F. Qi, G. Cao, D.-Y. Cho, and T. W. Noh, Phys. Rev. B **90**, 041105(R) (2014).

[15]    M. Hepting *et al.*, Phys. Rev. Lett. **113**, 227206 (2014).

[16] M. Hepting, D. Kukuruznyak, E. Benckiser, M. Le Tacon, and B. Keimer, Physica B **460**, 196 (2015).

[17] See Supplemental Material for further experimental and computational details.

[18] H. Gretarsson, N. H. Sung, M. Höppner, B. J. Kim, B. Keimer, and M. Le Tacon, Phys. Rev. Lett. **116**, 136401 (2016).

[19] N. H. Sung, H. Gretarsson, D. Proepper, J. Porras, M. Le Tacon, A. V. Boris, B. Keimer, and B. J. Kim, Philos. Mag. **96**, 413 (2016).

[20] H. Gretarsson, J. Sauceda, N. H. Sung, M. Höppner, M. Minola, B. J. Kim, B. Keimer, and M. Le Tacon, Phys. Rev. B **96**, 115138 (2017).

[21] J.-A. Yang, Y.-P. Huang, M. Hermele, T. Qi, G. Cao, and D. Reznik, Phys. Rev. B **91**, 195140 (2015).

[22] M. F. Cetin, P. Lemmens, V. Gnezdilov, D. Wulferding, D. Menzel, T. Takayama, K. Ohashi, and H. Takagi, Phys. Rev. B **85**, 195148 (2012).

[23] W. H. Weber and G. W. Ford, Phys. Rev. B **40**, 6890 (1989).

[24] P. A. Fleury and R. Loudon, Phys. Rev. **166**, 514 (1968).

[25] J. Kim *et al.*, Phys. Rev. Lett. **108**, 177003 (2012).

[26] S. A. Weidinger and W. Zwerger, Eur. Phys. J. B **88**, 237 (2015).

[27] Due to the large paramagnetic moments of Nd $f$-electrons and magnetic impurities, we could not obtain the magnetization data from $Sr_2IrO_4$/NGO and $Sr_2IrO_4$/LAO samples.

[28] A. M. Glazer, Acta Crystallogr. B **28**, 3384 (1972).

[29] From the resonant inelastic X-ray scattering experiment of Ref. [25], the second-nearest ($J_2 = -20$ meV) and third-nearest ($J_3 = 15$ meV) neighbor couplings are estimated an order of magnitude larger than the values of the DFT calculations. However, the two-magnon peak energy is determined by $J_1$ ($\omega_{2M} = 2.7J_1$) and the contributions from $J_2$ and $J_3$ are cancelled out according to a simple broken-bond counting argument ($\omega_{2M} = 3J_1 - 4J_2 - 4J_3$). See also Supplemental Material for details.

[30] M. Souri *et al.*, Phys. Rev. B **95**, 235125 (2017).

[31] K. B. Lyons, P. A. Fleury, J. P. Remeika, A. S. Cooper, and T. J. Negran, Phys. Rev. B **37**, 2353 (1988).

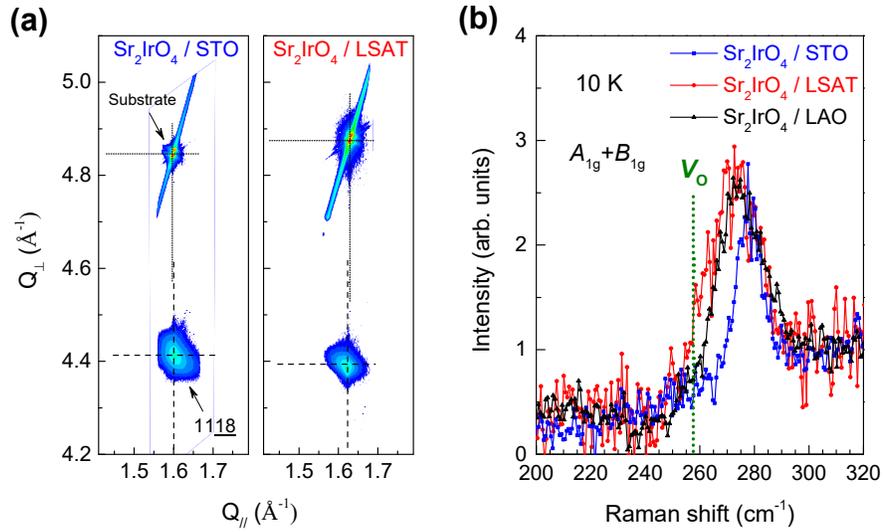

**FIG. 1. (a)** X-ray reciprocal space maps around the pseudo-cubic 103-reflection of STO and LSAT substrates. $Q_{//} \equiv 2\pi/a$ and $Q_{\perp} \equiv 2\pi/c$, where $a$ and $c$ are in-plane and out-of-plane lattice parameters, respectively. The 11$\underline{18}$-reflections from the $Sr_2IrO_4$ thin films are clearly visible. The average lattice parameters of the $Sr_2IrO_4$ thin films are obtained and listed in Table I. **(b)** Raman spectra of $Sr_2IrO_4$/STO, $Sr_2IrO_4$/LSAT, and $Sr_2IrO_4$/LAO taken at 10 K using the $z(xx)\bar{z}$ channel ($A_{1g}$ and $B_{1g}$). Oxygen deficient $Sr_2IrO_4$ is known to exhibit an oxygen-vacancy-related mode ($V_O$) at the Raman shift indicated by the dotted line (~258 cm$^{-1}$) [19], which is not observed in these spectra. The peaks at ~278 cm$^{-1}$ are an $A_{1g}$ phonon mode of $Sr_2IrO_4$ [20].

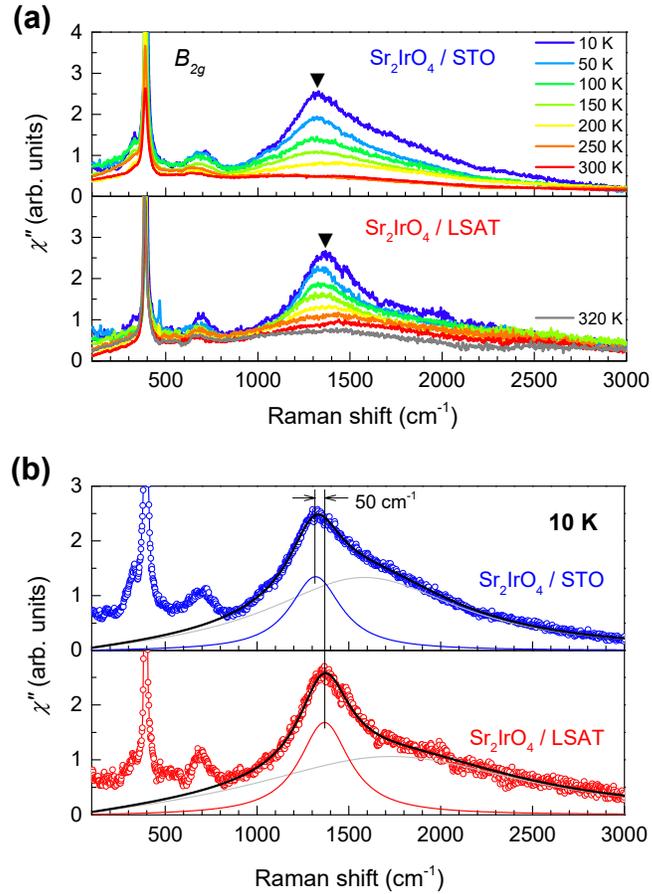

**FIG. 2.** (a) Temperature-dependent Raman spectra of $B_{2g}$ modes for $Sr_2IrO_4$/STO and $Sr_2IrO_4$/LSAT. The black triangles (▼) indicate the two-magnon peak energies at the lowest temperature (10 K). Sharp peaks at lower energies (< 800 cm$^{-1}$) are $B_{2g}$ phonons. (b) Raman spectra of $B_{2g}$ modes for $Sr_2IrO_4$/STO and $Sr_2IrO_4$/LSAT at 10 K. The solid lines are curve fits using two Lorentz oscillators for the broad, asymmetric two-magnon peaks above 900 cm$^{-1}$. The obtained two-magnon peak positions show a blueshift of ~50 cm$^{-1}$ from $Sr_2IrO_4$/STO to $Sr_2IrO_4$/LSAT.

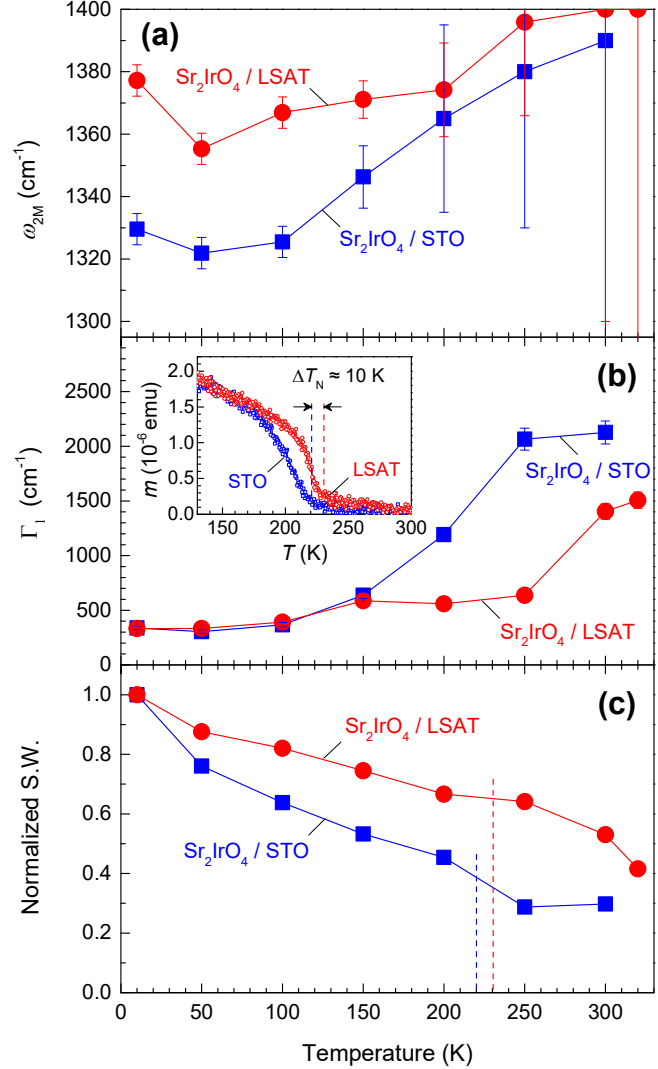

**FIG. 3. (a)** The two-magnon peak energies of $Sr_2IrO_4$/STO and $Sr_2IrO_4$/LSAT as a function of temperature. **(b)** The two-magnon peak widths of $Sr_2IrO_4$/STO and $Sr_2IrO_4$/LSAT as a function of temperature. The inset shows magnetizations vs. temperature measured with field-cooling at $H = 100$ Oe for the $Sr_2IrO_4$/STO and $Sr_2IrO_4$/LSAT samples. **(c)** The normalized spectral weights, *i.e.* the integrated intensities from 900 cm$^{-1}$ to 2500 cm$^{-1}$, of the two-magnon peaks as a function of temperature. The vertical dashed lines of blue and red color indicate $T_N$ of $Sr_2IrO_4$/STO and $Sr_2IrO_4$/LSAT samples, respectively, obtained from the inset of Fig. 3 (b).

# Supplemental Material

**Compressive-strain induced enhancement of exchange interactions and short-range magnetic order in $Sr_2IrO_4$ investigated by Raman spectroscopy**

A. Seo,[1,2] P. P. Stavropoulos,[3] H.-H. Kim,[2] K. Fürsich,[2] M. Souri,[1] J. G. Connell,[1] H. Gretarsson,[2] M. Minola,[2] H. Y. Kee,[3] and B. Keimer[2]

[1.] Department of Physics and Astronomy, University of Kentucky, 500 Rose St., Lexington, Kentucky 40506, USA
[2.] Max-Planck-Institut für Festkörperforschung, Heisenbergstraße 1, D-70569 Stuttgart, Germany
[3.] Department of Physics, University of Toronto, 60 St. George St., Toronto, Ontario, Canada M5S 1A7

## I. Computational details

For the electronic structure calculations and the atomic position optimization, we have employed the Vienna Ab-initio Simulation Package (VASP), which uses the projector-augmented wave (PAW) basis set [1, 2]. The plane-wave cutoff energy used is 500 eV and the k-point sampling used is a $9 \times 9 \times 9$ Monkhorst-Pack grid for the primitive unit cell. A revised Perdew-Burke-Ernzerhof generalized gradient approximation (PBEsol) is used for atomic position optimization and total-energy calculations [3, 4]. Optimization is performed with a force criterion of 1 meV/Å and without any symmetry constraints.

The lattice parameters are kept fixed due to the substrate and atomic position optimization is carried out for three cases: (1) without spin-orbit coupling (SOC) and on-site Coulomb interaction $U$, (2) including only SOC, and (3) including SOC and $U$ for the iridium sites. The on-site interactions are incorporated using Dudarev's rotationally invariant DFT+$U$ formalism [5] with the effective $U_{eff} = U - J_H = 2$ eV where $J_H$ is Hund's coupling, making the system magnetically ordered. For the sake of efficiency in structural relaxations, we have used the primitive unit cell which captures a week ferromagnetic order. Using structures with optimized atomic positions in each case above, the hopping parameters between the iridium $t_{2g}$ orbitals are computed by employing maximally localized Wannier orbital formalism (MLWF) [6, 7] implemented in the WANNIER90 package [8]. Owing to the layered structure of the material, a unit cell including only one layer of rotated octahedra is used for converging maximally localized Wannier orbitals.

## II. Results of atomic position optimization and Wannier orbitals

Optimized Ir-O-Ir bond-angles $\theta$ without SOC and $U$, hopping parameters and spin interactions are listed in Table SI (a). The substrate and the resulting $Sr_2IrO_4$ lattice parameters $a$ and $c$ from experiments label each strain case we examine. The angle $\theta$ after optimization is found to correlate predominantly with the variation of the lattice parameter $a$, which is consistent with layered nature of the material defining layers of corner sharing octahedra perpendicular to c.

The weak inter-layer coupling results in dispersion along $k_z$ being much less than that in the $k_x$, $k_y$ plane, which further justifies our isolated layer Wannier orbital calculation. We find hopping parameters between iridium $t_{2g}$ states of first nearest neighbors (1NN) and second nearest neighbors (2NN). The relevant spin model was derived in terms of $J_{eff} = 1/2$ moments in Ref. [9]. One can find hopping parameters between Jeff moments from the hopping parameters between $t_{2g}$ orbitals. The spin interactions for 1NN and 2NN can be computed from the following formulas:

$$J_1 = 4(t^2 - t_z^2)/\tilde{U}, \quad |D| = 8tt_z/\tilde{U}, \quad J_2 = 4t'^2/\tilde{U} \tag{1}$$

$$t = (t_{xz} + t_{yz} + t_{xy})/3, \quad t_z = (t_{yz,xz} - t_{xz,yz})/3, \quad t' = (t'_{xz} + t'_{yz} + t'_{xy})/3$$

An effective $\tilde{U} = 2$ eV was used to estimate the spin interactions. $t_{xz}$, $t_{yz}$, and $t_{xy}$ are the intra-orbital hoppings, $t_{yz,xz}/xz,yz$ is the inter-orbital hopping, and $t'$ refers to 2NN hopping. 1NN spin interactions consist of Heisenberg $J_1$ and Dzyaloshinskii-Moriya (DM) $D$ terms because of the absence of an inversion center along the 1NN bonds, however 2NN bonds do have inversion resulting in no 2NN DM term. More details on the microscopic equations above relating spin interaction parameters to hopping parameters and on-site electron interactions, can be found in Ref. [9].

Optimization of atomic positions including effects of SOC and on-site $U$, make for a heavier calculation, thus we choose the two extreme cases of the substrates, i.e. SrTiO$_3$ and NdGaO$_3$ to optimize atomic positions. Results when including only SOC are found in Table SI (b), and results when including both SOC and $U$ are found in Table SI (c). Comparing these results to the case without SOC and $U$ we find no appreciable difference between resulting angles $\theta$ and spin interactions.

**TABLE SI.** Table is split into three sections: results using lattice optimizations (a) without SOC and $U$, (b) only with SOC, and (c) with both SOC and $U$. The first column is a list of the substrates that Sr$_2$IrO$_4$ is grown on, and the second two are the imposed lattice parameters $a$ and $c$ used as input in *ab-initio* calculations. The resulting Ir-O-Ir bond angles $\theta$ after atomic position optimization, as well as hopping parameters between 1NN and 2NN $t_{2g}$ states computed from Wannier calculations, follow in the next three columns. The $t_{2g}$ hopping parameters are written in matrix form in the ($d_{xz}$, $d_{yz}$, $d_{xy}$) basis, with diagonal terms being $t_{xz}$, $t_{yz}$, $t_{xy}$ and nonzero off-diagonal terms being $t_{yz,xz/xz,yz}$. Using Eq. (1), the resulting $J_{eff}$ hopping parameters and spin interactions are given in the last six columns. All energy units are in units of meV and length units are in Å.

| substrate | $a$ | $c$ | $\theta$ | $t_{2g}$ hoppings 1NN | $t_{2g}$ hoppings 2NN | $t$ | $t_z$ | $t'$ | $J_1$ | $|D|$ | $J_2$ |
|---|---|---|---|---|---|---|---|---|---|---|---|
| (a) Optimized structures without SOC and $U$ ||||||||||||
| SrTiO$_3$ (100) | 5.5188 | 25.7022 | 154.74 | $\begin{pmatrix}-205 & -29 & 0\\ 29 & -54 & 0\\ 0 & 0 & -212\end{pmatrix}$ | $\begin{pmatrix}4 & -4 & 0\\ -4 & 44 & 0\\ 0 & 0 & -122\end{pmatrix}$ | -157 | -19 | -25 | 49 | 12 | 1.2 |
| LSAT (100) | 5.5129 | 25.7996 | 154.36 | $\begin{pmatrix}-211 & -25 & 0\\ 25 & -55 & 0\\ 0 & 0 & -210\end{pmatrix}$ | $\begin{pmatrix}4 & -5 & 0\\ -5 & 44 & 0\\ 0 & 0 & -122\end{pmatrix}$ | -159 | -17 | -25 | 50 | 11 | 1.2 |
| NdGaO$_3$ (110) | 5.3979 | 25.9004 | 151.76 | $\begin{pmatrix}-269 & -33 & 0\\ 33 & -55 & 0\\ 0 & 0 & -206\end{pmatrix}$ | $\begin{pmatrix}3 & -4 & 0\\ -4 & 45 & 0\\ 0 & 0 & -129\end{pmatrix}$ | -177 | -22 | -27 | 61 | 16 | 1.5 |
| LaAlO$_3$ (110) | 5.4324 | 25.9709 | 152.33 | $\begin{pmatrix}-254 & -30 & 0\\ 30 & -54 & 0\\ 0 & 0 & -205\end{pmatrix}$ | $\begin{pmatrix}3 & -4 & 0\\ -4 & 44 & 0\\ 0 & 0 & -126\end{pmatrix}$ | -171 | -20 | -26 | 58 | 14 | 1.4 |
| (b) Optimized structures including SOC ||||||||||||
| SrTiO$_3$ (100) | 5.5188 | 25.7022 | 155.50 | $\begin{pmatrix}-211 & -27 & 0\\ 27 & -54 & 0\\ 0 & 0 & -219\end{pmatrix}$ | $\begin{pmatrix}5 & -5 & 0\\ -5 & 44 & 0\\ 0 & 0 & -124\end{pmatrix}$ | -161 | -18 | -25 | 51 | 12 | 1.3 |
| NdGaO$_3$ (110) | 5.3979 | 25.9004 | 152.15 | $\begin{pmatrix}-274 & -32 & 0\\ 32 & -55 & 0\\ 0 & 0 & -208\end{pmatrix}$ | $\begin{pmatrix}3 & -5 & 0\\ -5 & 44 & 0\\ 0 & 0 & -127\end{pmatrix}$ | -179 | -21 | -27 | 63 | 15 | 1.4 |
| (c) Optimized structures including SOC + $U$ ||||||||||||
| SrTiO$_3$ (100) | 5.5188 | 25.7022 | 155.29 | $\begin{pmatrix}-206 & -26 & 0\\ 26 & -55 & 0\\ 0 & 0 & -217\end{pmatrix}$ | $\begin{pmatrix}5 & -5 & 0\\ -5 & 44 & 0\\ 0 & 0 & -123\end{pmatrix}$ | -159 | -17 | -25 | 50 | 11 | 1.2 |
| NdGaO$_3$ (110) | 5.3979 | 25.9004 | 152.08 | $\begin{pmatrix}-271 & -32 & 0\\ 32 & -55 & 0\\ 0 & 0 & -207\end{pmatrix}$ | $\begin{pmatrix}3 & -4 & 0\\ -4 & 45 & 0\\ 0 & 0 & -127\end{pmatrix}$ | -178 | -21 | -26 | 62 | 15 | 1.4 |

## III. X-ray reciprocal space maps of Sr$_2$IrO$_4$/NGO and Sr$_2$IrO$_4$/LAO

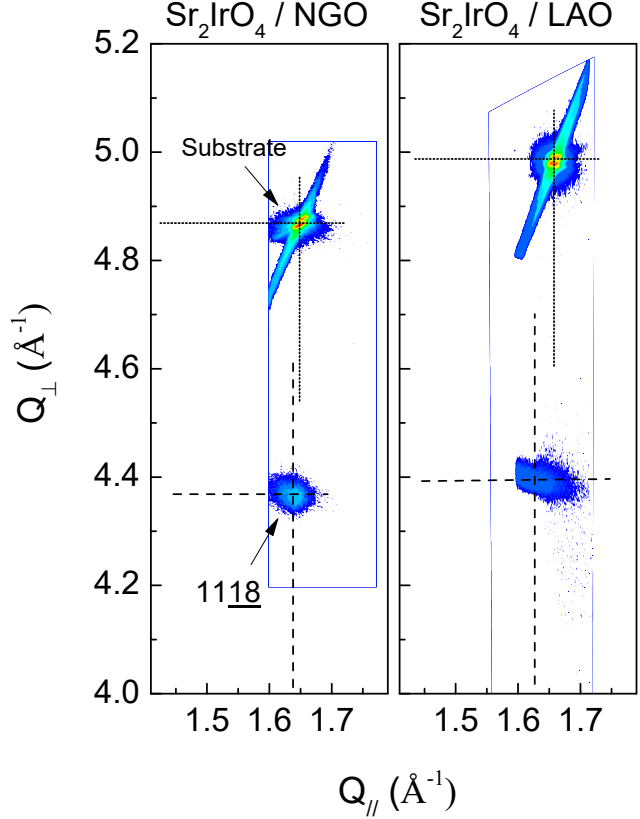

**FIG S1.** X-ray reciprocal space maps near the pseudo-cubic 103-reflection of NGO and LAO substrates. $Q_{//} \equiv 2\pi/a$ and $Q_{\perp} \equiv 2\pi/c$, where $a$ and $c$ are in-plane and out-of-plane lattice parameters, respectively. The 11$\underline{18}$-reflections from the Sr$_2$IrO$_4$ thin films are clearly visible. However, there are relatively large lattice relaxation (See the difference between vertical dotted lines (substrates) and dashed lines (thin films)) presumably due to large lattice mismatch compared to the thin films grown on STO and LSAT substrates. The average lattice parameters of the Sr$_2$IrO$_4$ thin films are listed in Table SII.

**Table SII.** Lattice parameters and misfit strain values of Sr$_2$IrO$_4$/NGO and Sr$_2$IrO$_4$/LAO.

| Samples | $a$ (Å) | $c/2$ (Å) | Misfit strain tensile (+) / compressive (−) | |
| --- | --- | --- | --- | --- |
| | | | In-plane (%) | Out-of-plane (%) |
| Sr$_2$IrO$_4$/NGO | 3.81 | 13.01 | −1.9 (±0.3) | +0.91 (±0.2) |
| Sr$_2$IrO$_4$/LAO | 3.84 | 12.94 | −1.2 (±0.5) | +0.37 (±0.2) |

## IV. Extraction of Raman spectra of Sr$_2$IrO$_4$ thin films

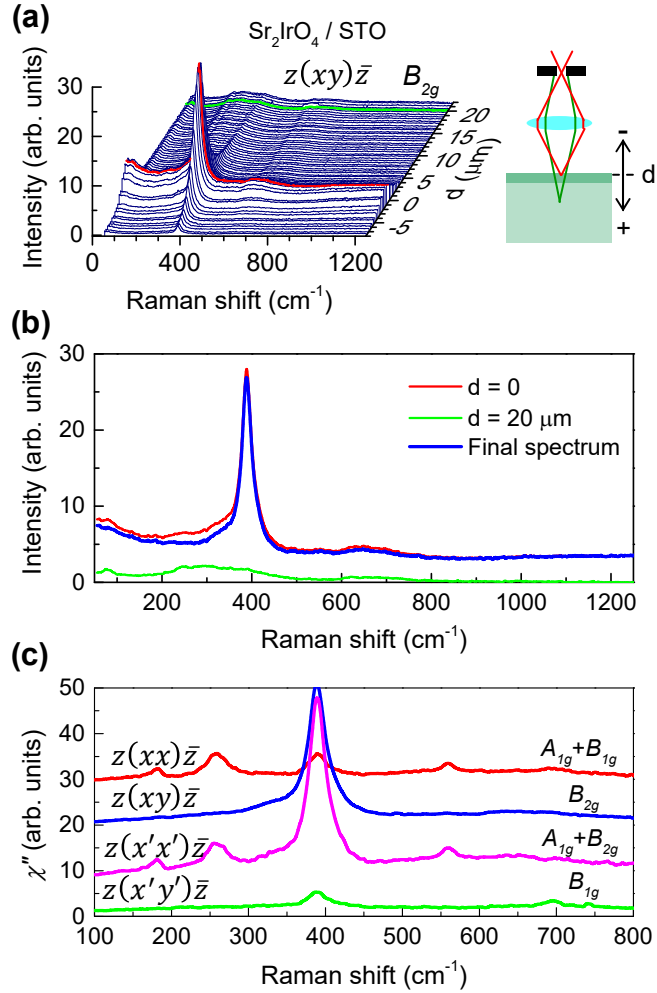

**FIG S2. (a)** Beam-position ($d$) dependent Raman spectra taken at room temperature for Sr$_2$IrO$_4$/STO in confocal geometry, as shown schematically on the right. The spectra at $d = 0$ and $d = 20$ μm are used for obtaining the spectrum of the thin film, as shown in **(b)**. **(c)** Polarization-dependent Raman spectra of Sr$_2$IrO$_4$/STO taken at room temperature, which show consistent phonon modes and selection rules with an Sr$_2$IrO$_4$ single crystal [10].

## V. Temperature-dependent Raman spectra of $Sr_2IrO_4$/NGO and $Sr_2IrO_4$/LAO

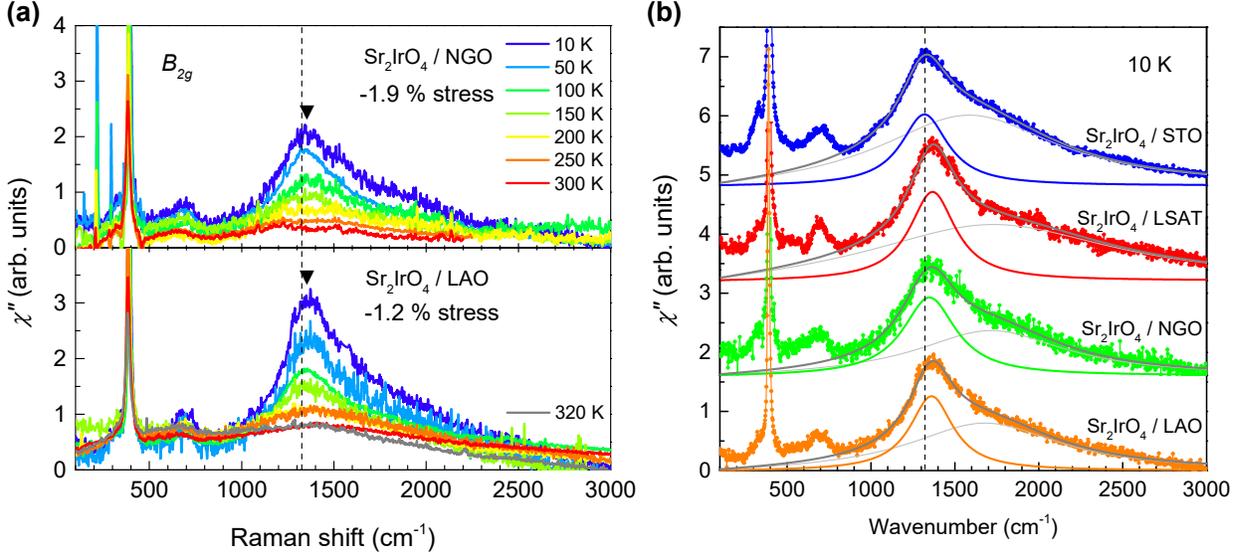

**FIG S3. (a)** Temperature-dependent Raman spectra of $B_{2g}$ modes for $Sr_2IrO_4$/NGO and $Sr_2IrO_4$/LAO, which are under volume-averaged compressive strain of -1.9 % and -1.2 %, respectively. Clear two-magnon scattering centered at ~1300 cm$^{-1}$ is observed. The black triangles (▼) indicate two-magnon peak energies at the lowest temperature (10 K). Sharp peaks at lower energies (< 800 cm$^{-1}$) are $B_{2g}$ phonons of $Sr_2IrO_4$ (and some features from the substrates). The vertical dashed line is drawn from the two-magnon peak energy of $Sr_2IrO_4$/STO to compare with these samples. Similarly to $Sr_2IrO_4$/LSAT, these $Sr_2IrO_4$/NGO and $Sr_2IrO_4$/LAO samples show higher two-magnon peak energies than $Sr_2IrO_4$/STO by about 50 cm$^{-1}$ although there is relatively large strain relaxation, as shown in Fig. S1. **(b)** Raman spectra of $B_{2g}$ modes at 10 K and curve fits (solid lines) using two Lorentz oscillators for the broad, asymmetric peaks of two-magnon scattering above 900 cm$^{-1}$ for $Sr_2IrO_4$/STO, $Sr_2IrO_4$/LSAT, $Sr_2IrO_4$/NGO, and $Sr_2IrO_4$/LAO. The vertical dashed line is drawn from the two-magnon peak energy of $Sr_2IrO_4$/STO to compare with other samples. The obtained two-magnon peak positions show the blueshift of ~50 cm$^{-1}$ from $Sr_2IrO_4$/STO to other $Sr_2IrO_4$ thin films with compressive strain.

## VI. Spectral weights of two-magnon Raman scattering

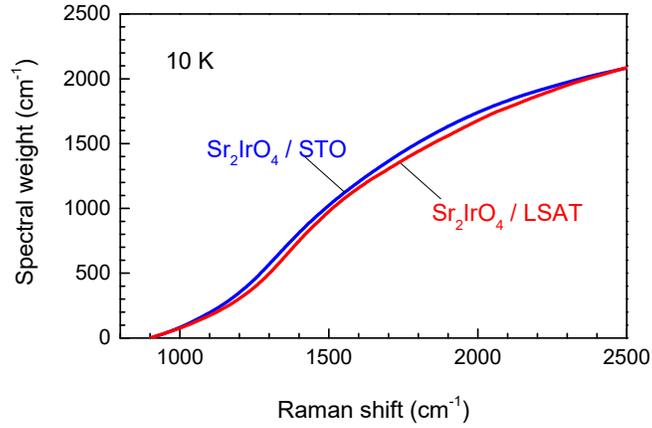

**FIG S4.** The spectral weight, *i.e.* the integrated intensity, of two-magnon Raman scattering at 10 K as a function of Raman shift between 900 cm$^{-1}$ and 2500 cm$^{-1}$. The result shows that Sr$_2$IrO$_4$/LSAT has overall spectral weight at higher energies than Sr$_2$IrO$_4$/STO, which is consistent with the blueshift of two-magnon peak energies shown in Fig. 2 and Fig. S3 (b).

# References

[1] G. Kresse and J. Furthmüller, Phys. Rev. B **54**, 11169 (1996).

[2] G. Kresse and D. Joubert, Phys. Rev. B **59**, 1758 (1999).

[3] J. P. Perdew, A. Ruzsinszky, G. I. Csonka, O. A. Vydrov, G. E. Scuseria, L. A. Constantin, X. Zhou, and K. Burke, Phys. Rev. Lett. **100**, 136406 (2008).

[4] J. P. Perdew, A. Ruzsinszky, G. I. Csonka, O. A. Vydrov, G. E. Scuseria, L. A. Constantin, X. Zhou, and K. Burke, Phys. Rev. Lett. **102**, 039902 (2009).

[5] S. L. Dudarev, G. A. Botton, S. Y. Savrasov, C. J. Humphreys, and A. P. Sutton, Phys. Rev. B **57**, 1505 (1998).

[6] N. Marzari and D. Vanderbilt, Phys. Rev. B **56**, 12847 (1997).

[7] I. Souza, N. Marzari, and D. Vanderbilt, Phys. Rev. B **65**, 035109 (2001).

[8] A. A. Mostofi, J. R. Yates, Y.-S. Lee, I. Souza, D. Vanderbilt, and N. Marzari, Comput. Phys. Commun. **178**, 685 (2008).

[9] J.-M. Carter, V. Shankar V., and H.-Y. Kee, Phys. Rev. B **88**, 035111 (2013).

[10] H. Gretarsson, J. Sauceda, N. H. Sung, M. Höppner, M. Minola, B. J. Kim, B. Keimer, and M. Le Tacon, Phys. Rev. B **96**, 115138 (2017).